\documentclass[12pt,titlepage]{article}
\usepackage{amsmath}
\usepackage{amsfonts}
\usepackage{charter}
\usepackage{geometry}
\usepackage[doublespacing]{setspace}
\usepackage{mathrsfs}
\usepackage{rotating}

\newtheorem{theorem}{Theorem}

\newtheorem{proposition}[theorem]{Proposition}

\input{tcilatex}
\renewcommand{\mathcal}{\mathscr}

\begin{document}

\title{Outlier detection and trimmed estimation for general functional data}
\author{Daniel Gervini \\
%EndAName
\emph{Department of Mathematical Sciences}\\
\emph{University of Wisconsin--Milwaukee}\\
\emph{P.O. Box 413, Milwaukee, WI 53201}\\
\textsf{gervini@uwm.edu}}
\maketitle

\begin{abstract}
This article introduces trimmed estimators for the mean and covariance
function of general functional data. The estimators are based on a new
measure of \textquotedblleft outlyingness\textquotedblright\ or data depth\
that is well defined on any metric space, although this paper focuses on
Euclidean spaces. We compute the breakdown point of the estimators and show
that the optimal breakdown point is attainable for the appropriate choice of
tuning parameters. The small-sample behavior of the estimators is studied by
simulation, and we show that they have better outlier-resistance properties
than alternative estimators. This is confirmed by two real-data
applications, that also show that the outlyingness measure can be used as a
graphical outlier-detection tool in functional spaces where visual screening
of the data is difficult.

\emph{Key Words:} Breakdown Point; Data Depth; Robust Statistics; Stochastic
Processes.
\end{abstract}

\section{Introduction}

Many statistical applications today involve data that does not fit into the
classical univariate or multivariate frameworks; for example, growth curves,
spectral curves, and time-dependent gene expression profiles. These are
samples of functions, rather than numbers or vectors. We can think of them
as realizations of a stochastic process with sample paths in $\mathcal{L}%
^{2}(\mathbb{R})$, the space of square-integrable functions. The statistical
analysis of function-valued data has received a lot of attention in recent
years (see e.g.~Ramsay and Silverman 2002, 2005, and references therein).
However, most of the work on Functional Data Analysis has focused on
univariate curves; in many applications, the sample functions are not
univariate.

\FRAME{ftbpFU}{5.5019in}{3.9954in}{0pt}{\Qcb{Excitation--Emission Matrices.
Four samples of log-EEMs.}}{\Qlb{fig:EEM_samples}}{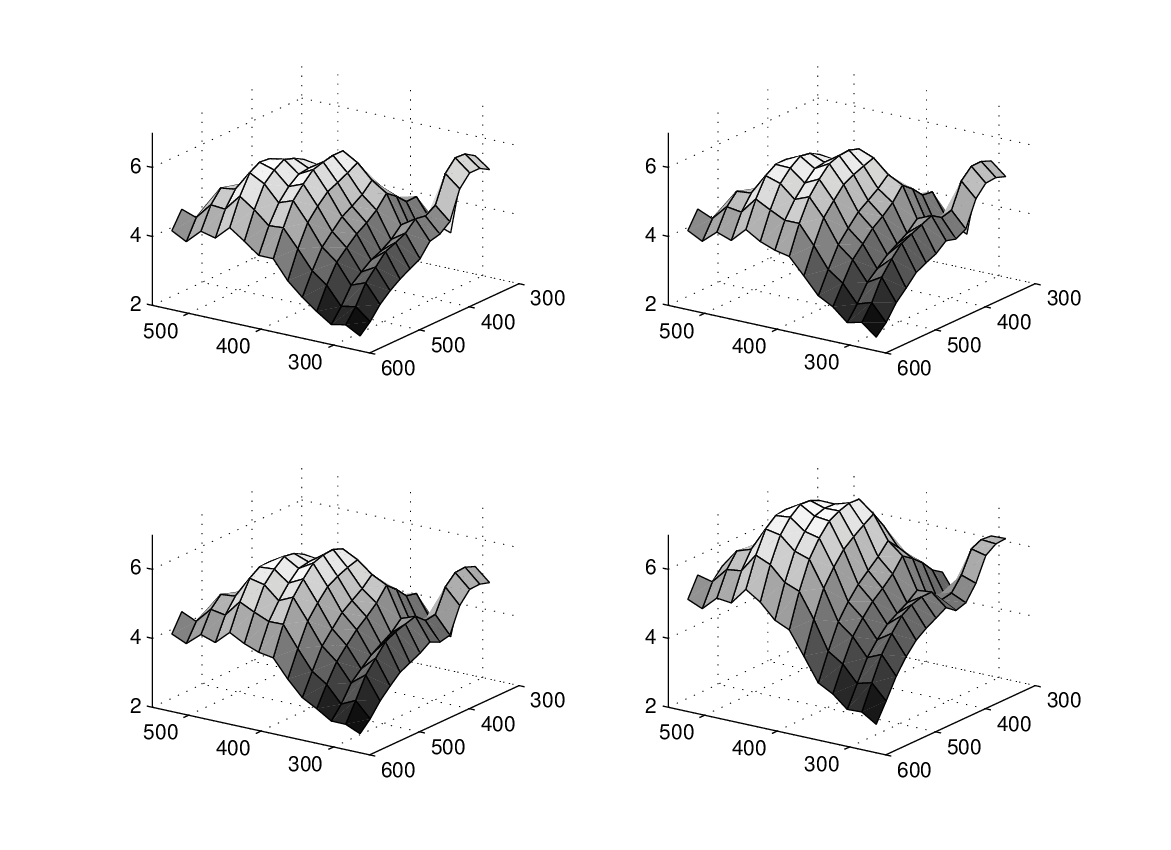}{\special%
{language "Scientific Word";type "GRAPHIC";maintain-aspect-ratio
TRUE;display "USEDEF";valid_file "F";width 5.5019in;height 3.9954in;depth
0pt;original-width 7.8075in;original-height 5.6585in;cropleft "0";croptop
"1";cropright "1";cropbottom "0";filename 'samples.eps';file-properties
"XNPEU";}}

Consider, for example, excitation-emission matrices (EEMs), which are common
in Chemometrics. When certain fluorescent substances are exposed to light of
wavelength $s$, they emit light at wavelength $t$. The resulting light
intensity $X$ is then a bivariate function $X(s,t)$, that is, a $\mathbb{R}%
^{2}\rightarrow \mathbb{R}$ function. Mortensen and Bro (2006) analyzed a
collection of 338 such surfaces; the logarithms of four of them are shown in
Figure \ref{fig:EEM_samples}. A movie showing the 338 surfaces in quick
succession is available on the author's website; it is clear in this movie
that there are some atypical surfaces in the data set. For example, Figure %
\ref{fig:EEM_samples} shows that even after taking logarithms, the surface
on the lower right corner is out of line compared to the other three. These
atypical objects is what we will refer to as \textquotedblleft
outliers\textquotedblright\ in this paper; that is, objects that depart from
the main modes of variability of the majority of the data. Note that since
each functional object typically consists of many measurements taken at
different time points (or wavelength points, in this case), a few of those
measurements could be outlying without the whole surface being necessarily
atypical. But that kind of isolated measurement errors are not the type of
outliers we are interested in in this paper; they have been addressed in the
robust smoothing literature (e.g.~Shi and Li 1995, Jiang and Mack 2001.)

\FRAME{ftbpFU}{3.5198in}{2.6498in}{0pt}{\Qcb{Handwritten Digits Example.
Eight samples of the number \textquotedblleft five\textquotedblright .}}{%
\Qlb{fig:digit_samples}}{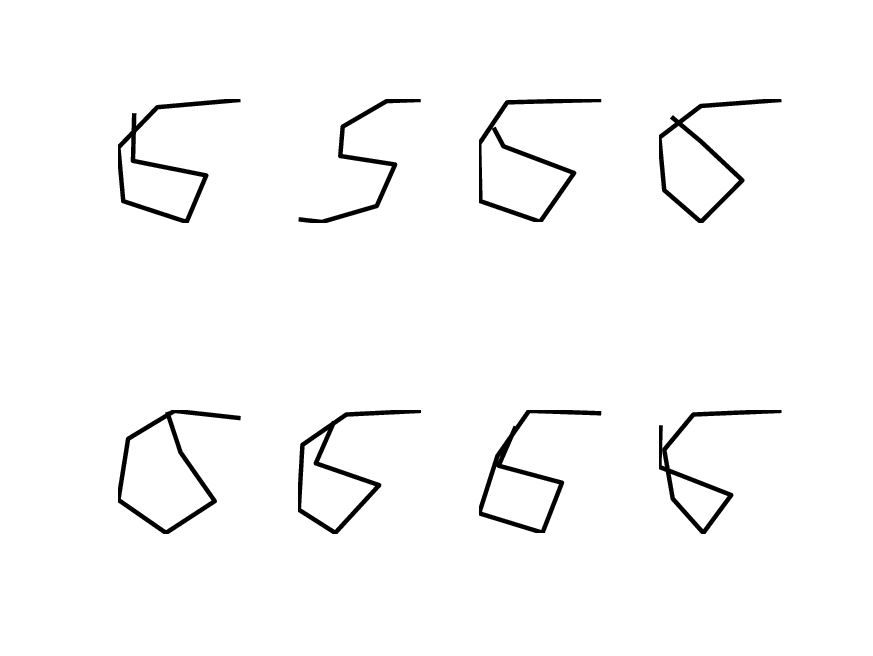}{\special{language "Scientific
Word";type "GRAPHIC";maintain-aspect-ratio TRUE;display "USEDEF";valid_file
"F";width 3.5198in;height 2.6498in;depth 0pt;original-width
5.8219in;original-height 4.3708in;cropleft "0";croptop "1";cropright
"1";cropbottom "0";filename 'digit_samples.eps';file-properties "XNPEU";}}

As a second example, consider a digit recognition problem. Eight handwritten
\textquotedblleft fives\textquotedblright , from a total of 1055 samples,
are shown in Figure \ref{fig:digit_samples}. The planar trajectory of the
pen tip is a curve $(x(t),y(t))$ in $\mathbb{R}^{2}$, where the variable $t$
is time, so the digits are $\mathbb{R}\rightarrow \mathbb{R}^{2}$ functions
(note that we are ignoring a third variable, $z(t)$, the distance between
the pen tip and the writing pad). In Figure \ref{fig:digit_samples} we see
that some of the handwritten digits look more like \textquotedblleft
sixes\textquotedblright\ than \textquotedblleft fives\textquotedblright .
The reason for this is going to be explained in Section \ref{sec:Examples},
but it is clear at this point that the sample of \textquotedblleft
fives\textquotedblright\ is not homogeneous; it contains either isolated
outliers or systematic clusters that are hard to guess a priori.

These examples show two things, which are the main motivation for this
paper: \emph{(i)} functional data belonging to spaces more complicated than $%
\mathcal{L}^{2}(\mathbb{R})$ are encountered in practice, and \emph{(ii)}
outliers may be present in a sample but, due to the complexity of the data,
visual screening of the data set may be impractical or impossible. The
problem of robust estimation in functional spaces has been addressed by some
authors, including Locantore et al.~(1999), Fraiman and Muniz (2001), Cuevas
et al.~(2007), Gervini (2008), and L\'{o}pez-Pintado and Romo (2009). But
all of these papers deal with univariate curves. Some of these methods can
be extended to more complex spaces in a more or less straightforward way,
but some of them cannot. For example, the methods of Fraiman and Muniz
(2001) and L\'{o}pez-Pintado and Romo (2009) are based on data-depth notions
that require an ordering of the response variables and then they cannot be
extended to vector-valued functions like the handwritten digits in an
obvious way. On the other hand, the projection-based methods discussed in
Cuevas et al.~(2007) and the spatial median and the spherical principal
components of Locantore et al.~(1999) and Gervini (2008) can be extended to
any Euclidean space; but Gervini (2008) found that the breakdown point of
the spherical principal components is very low, so a third goal of this
paper is to develope principal component estimators that are more robust
than the spherical principal components but not as computationally demanding
as the projection-based methods of Cuevas et al.~(2007).

The estimators introduced in this article are based on a measure of
\textquotedblleft outlyingness\textquotedblright\ that can be defined on any
metric space, but we will restrict ourselves to Euclidean spaces, where
principal components can also be defined. These estimators are easy to
compute and turned out to have very good robustness properties. We prove in
Section \ref{sec:Properties} that they can attain the optimal 50\% breakdown
point (i.e.~they can resist up to 50\% of outliers in the data). In our
simulation study (Section \ref{sec:Simulations}) they outperformed most of
the alternative estimators cited above. The paper also studies other
theoretical properties in Section \ref{sec:Properties}, and analyzes in more
detail the two applications mentioned above (Section \ref{sec:Examples}).
Proofs of the theoretical results and an additional real-data application
can be found in a technical supplement available on the author's webpage.

\section{\label{sec:Definition}Trimmed estimators based on interdistances}

\subsection{A measure of \textquotedblleft outlyingness\textquotedblright}

Let $\{X_{1},\ldots ,X_{n}\}$ be a sample in a Euclidean space $\mathcal{H}$%
, i.e.~a linear space endowed with an inner product $\langle \cdot ,\cdot
\rangle $ (for instance, $\mathcal{L}^{2}(\mathbb{R})$ with its canonical
inner product $\langle f,g\rangle =\int fg$.) The inner product induces the
norm $\Vert f\Vert =\langle f,f\rangle ^{1/2}$ in $\mathcal{H}$, and this
norm induces the distance function $d(f,g)=\Vert f-g\Vert $, so any
Euclidean space is a metric space. Let us consider the set of interdistances 
$\{d(X_{i},X_{j})\}$. An observation $X_{i}$ can be seen as an outlier if
it's far from \emph{most} of the other observations (not necessarily from 
\emph{all} of them, because outliers sometimes form clusters). Given $\alpha
\in \lbrack 0,1]$, we define the $\alpha $-radius $r_{i}$ as the distance
between $X_{i}$ and the $\lceil \alpha n\rceil $-th closest observation,
where $\lceil x\rceil $ denotes the integer closest to $x$ from above. This
is the radius of the smallest ball centered at $X_{i}$ that covers $%
100\alpha \%$ of the observations. Intuitively, $r_{i}$ will be small where
the data is dense and large where the data is sparse (see Proposition \ref%
{prop:r_depth} in Section \ref{sec:Properties}). Therefore, the rank of $%
r_{i}$ in the set $\{r_{1},\ldots ,r_{n}\}$ will be a measure of the
\textquotedblleft outlyingness\textquotedblright\ of $X_{i}$: the more
isolated $X_{i}$ is, the larger $r_{i}$ will be compared to the other radii.

In principle the coverage parameter $\alpha $ could be any number between 0
and 1, but note that if there is a tight cluster of $n^{\ast }$ outliers and 
$\lceil \alpha n\rceil <n^{\ast }$, then the radii of the outliers will be
small, perhaps even smaller than the radii of the \textquotedblleft
good\textquotedblright\ observations, which would render them useless for
our purposes. Therefore $\alpha $ must be large enough that at least one
good observation is captured by $r_{i}$ whenever $X_{i}$ is an outlier.
Since $n^{\ast }$ can be as large as $n/2$, in general only $\alpha \geq .50$
will guarantee this. On the other hand, taking $\alpha >.50$ may cause the
opposite problem: that an outlying observation will always be captured by $%
r_{i}$ when $X_{i}$ is not an outlier, making the radii of the
\textquotedblleft good\textquotedblright\ observations too large (the
formalization of these heuristics constitute the proof of Proposition \ref%
{prop:BDP} in Section \ref{sec:Properties}). For these reasons we will
always take $\alpha =.50$ for estimation purposes. However, for
outlier-screening purposes it is instructive to see boxplots and histograms
of the radii for values of $\alpha $ less than $.50$; the outliers tend to
emerge clearly and consistently as $\alpha $ increases.

At this point some comments about the actual computation of the
interdistances are in order. First, note that all the interdistances can be
computed from the set of inner products $\{\langle X_{i},X_{j}\rangle \}$,
since $d^{2}(X_{i},X_{j})=\langle X_{i},X_{i}\rangle +\langle
X_{j},X_{j}\rangle -2\langle X_{i},X_{j}\rangle $. It is easy to compute the
pairwise inner products when the sample objects have been pre-smoothed, or
even if they have not been pre-smoothed but they were sampled on a regular
common grid without much random error. In that case, a basic numerical
integration method such as the trapezoidal rule will give accurate results
(see Gervini 2008, Theorem 1). But if the $X_{i}$s were sampled on sparse
and irregular grids, perhaps with a different grid for each individual, then
it will not be possible to estimate all pairwise inner products and this
method cannot be applied (the other methods mentioned in the introduction
cannot be applied either, since they are based on pre-smoothed data).

\subsection{Trimmed estimators}

In addition to being useful outlying-screening tools, the radii can be used
to construct robust estimators of the mean, the covariance function and the
principal components of the process under consideration. For a stochastic
process $X$ in $\mathcal{H}$ with $E(\Vert X\Vert ^{2})<\infty $, the mean
operator $\mathfrak{M}$ and the covariance operator $\mathfrak{C}$ are
defined as follows: $\mathfrak{M}:\mathcal{H}\rightarrow \mathbb{R}$ is
given by $\mathfrak{M}f=E(\langle f,X\rangle )$, and $\mathfrak{C}:\mathcal{H%
}\times \mathcal{H}\rightarrow \mathbb{R}$ is given by $\mathfrak{C}(f,g)=%
\mathrm{cov}(\langle f,X\rangle ,\langle g,X\rangle )$ (these quantities are
well defined because $\langle f,X\rangle $ and $\langle g,X\rangle $ are
real-valued random variables with finite variances for any $f$ and $g$ in $%
\mathcal{H}$.) By Riesz Representation Theorem there exists a unique $\mu
\in \mathcal{H}$ such that $\mathfrak{M}f=\langle f,\mu \rangle $, which we
call $E(X)$ (this is one way to define the expectation of a stochastic
process in a Euclidean space.)

In a Euclidean space it is also possible to define principal directions of
variability, or principal components. The first principal component of $X$
is $\phi _{1}\in \mathcal{H}$ that maximizes $\mathrm{var}(\langle
f,X\rangle )$ among $f\in \mathcal{H}$ with $\Vert f\Vert =1$; the second
principal component is $\phi _{2}\in \mathcal{H}$ that maximizes $\mathrm{var%
}(\langle f,X\rangle )$ among $f\in \mathcal{H}$ with $\Vert f\Vert =1$ and $%
\langle f,\phi _{1}\rangle =0$; and so on. It can be shown (Gohberg et
al.~2003, chap.~IV) that the principal components are eigenfunctions of the
covariance operator and they are countable; that is, $\mathfrak{C}(\phi
_{k},\cdot )=\lambda _{k}\langle \phi _{k},\cdot \rangle $ with $\lambda
_{k}\in \mathbb{R}$ and $\lambda _{k}\geq 0$.

The classical estimators of these quantities (the sample mean, covariance,
and principal components) are not resistant to outliers. As a more robust
alternative we propose trimmed estimators based on the radii. Specifically,
given a trimming proportion $\beta \in \lbrack 0,.50]$ we define $w(X_{i})=%
\mathbb{I}\{r_{i}<r_{(\lceil (1-\beta )n\rceil )}\}$ and 
\begin{equation}
\hat{\mu}=\frac{1}{\sum_{i=1}^{n}w(X_{i})}\sum_{i=1}^{n}w(X_{i})X_{i},
\label{eq:mu_hat_gen}
\end{equation}%
\begin{equation}
\widehat{\mathfrak{C}}(f,g)=\frac{1}{\sum_{i=1}^{n}w(X_{i})}%
\sum_{i=1}^{n}w(X_{i})\langle X_{i}-\hat{\mu},f\rangle \langle X_{i}-\hat{\mu%
},g\rangle .  \label{eq:C_hat_gen}
\end{equation}%
These are \textquotedblleft hard-trimmed\textquotedblright\ estimators,
where a 0-1 weight function is used. More generally, we can define weights
of the form $w(X_{i})=g(\mathrm{rank}(r_{i})/n)$, where $g:[0,1]\rightarrow 
\mathbb{R}^{+}$ is a bounded, non-negative and non-increasing function such
that $g(t)>0$ for $t<1-\beta $ and $g(t)=0$ for $t\geq 1-\beta $.
\textquotedblleft Soft-trimming\textquotedblright\ weights are obtained with
a smooth function $g$ such as 
\begin{equation}
g(r)=\left\{ 
\begin{array}{lll}
1, &  & 0\leq r\leq a, \\ 
(r-b)\left[ \frac{1}{(a-b)}+\frac{(r-a)\{2r-(a+b)\}}{(b-a)^{3}}\right] , & 
& a\leq r\leq b, \\ 
0, &  & r\geq b,%
\end{array}%
\right.  \label{eq:weight_function_g}
\end{equation}%
where $a=1-\beta _{1}$ for some $\beta _{1}>\beta $, and $b=1-\beta $. This
function downweights the largest $100\beta _{1}\%$ radii, and cuts off the
largest $100\beta \%$ radii completely; we can take, for example, $\beta
_{1}=.50$ and $\beta =.20$.

Trimmed estimators based on various measures of data depth have been
proposed in other contexts, in particular in multivariate analysis (Fraiman
and Meloche 1999, Liu et al.~1999, Serfling 2006, Zuo and Serfling 2000, Zuo
et al.~2004). The behavior of these estimators varies according to the
specific data-depth measure that is being used, but as a general rule, their
outlier resistance increases as $\beta $ increases and their efficiency
decreases as $\beta $ increases (see e.g.~Stigler 1973; Van der Vaart 1998,
chap.~22; Maronna et al.~2006, chap.~2). Since there is a trade-off between
robustness and efficiency, we recommend choosing $\beta $ in a data-driven
way: a histogram of the radii usually gives a good idea of the proportion of
outliers in the sample, and this value could be used as $\beta $. A more
objective alternative, suggested by a referee, is to fit a mixture of two
Gamma distributions to the sample of radii and take as $\beta $ the
proportion of observations in the smaller group. If instead of these
data-driven choices of $\beta $ the user prefers to use a fixed $\beta $,
our simulations showed that \textquotedblleft
soft-trimming\textquotedblright\ weights like (\ref{eq:weight_function_g})
are preferrable to \textquotedblleft hard-trimming\textquotedblright\
weights (see Section \ref{sec:Simulations}).

Just like the radii (and therefore the weights $w(X_{i})$) depend on the
data only through the inner products $\{\langle X_{i},X_{j}\rangle \}$ as
mentioned in the previous section, the principal components of (\ref%
{eq:C_hat_gen}) can also be computed entirely from the inner products, as
explained in Gervini (2008) and Jolliffe (2002, ch.~3.5): if $\tilde{w}%
_{i}=w(X_{i})/\sum_{i=1}^{n}w(X_{i})$, then $\hat{\phi}_{k}=%
\sum_{i=1}^{n}(c_{ki}/l_{k}^{1/2})\tilde{w}_{i}^{1/2}(X_{i}-\hat{\mu})$ and $%
\hat{\lambda}_{k}=l_{k}$, where $\mathbf{c}_{k}$ is the $k$th unit-norm
eigenvector of the matrix $\mathbf{G}\in \mathbb{R}^{n\times n}$ with
elements $G_{ij}=\langle \tilde{w}_{i}^{1/2}(X_{i}-\hat{\mu}),\tilde{w}%
_{j}^{1/2}(X_{j}-\hat{\mu})\rangle $, and $l_{k}$ is the $k$th eigenvalue
(the $G_{ij}$s can be expressed entirely in terms of the $\langle
X_{i},X_{j}\rangle $s and the $\tilde{w}_{i}$s, after some algebra). The
applicability of these estimators will then be limited only by the
possibility of computing all pairwise inner products. As mentioned before,
this is generally not possible if the data objects were sparsely and
irregularily sampled, and alternative estimation methods must be sought. For
instance, the reduced-rank $t$-model estimators of Gervini (2010), which
were originally developed for sparsely sampled univariate curves, can be
extended to more general functional spaces, but this is clearly outside the
scope of this paper.

\section{\label{sec:Properties}Properties of the estimators}

\subsection{Finite-sample properties}

Location and scatter estimators must satisfy certain equivariance
properties, in order to be proper measures of \textquotedblleft
location\textquotedblright\ and \textquotedblleft scatter\textquotedblright
. A location estimator must be translation equivariant: if $\hat{\mu}$ is
the estimator based on the sample $\{X_{1},\ldots ,X_{n}\}$, then the
estimator based on the sample $\{X_{1}+c,\ldots ,X_{n}+c\}$, with $c\in 
\mathcal{H}$, must be $\hat{\mu}+c$. Other desirable properties are scale
and rotation equivariance: if $\hat{\mu}$ is the estimator based on the
sample $\{X_{1},\ldots ,X_{n}\}$, then the estimator based on the sample $\{a%
\mathfrak{U}X_{1},\ldots ,a\mathfrak{U}X_{n}\}$, with $\mathfrak{U}$ a
unitary operator and $a\in \mathbb{R}$, must be $a\mathfrak{U}\hat{\mu}$ (a
unitary operator is $\mathfrak{U}:\mathcal{H}\rightarrow \mathcal{H}$ such
that $\Vert \mathfrak{U}f\Vert =\Vert f\Vert $ for every $f\in \mathcal{H}$%
.) A scatter estimator, on the other hand, must be translation invariant
(i.e.~remain unchanged under translations) and rotation and scale
equivariant in the following sense: if $\widehat{\mathfrak{C}}(\cdot ,\cdot
) $ is the covariance estimator based on the sample $\{X_{1},\ldots ,X_{n}\}$%
, then the covariance estimator based on the sample $\{a\mathfrak{U}%
X_{1},\ldots ,a\mathfrak{U}X_{n}\}$ must be $a^{2}\widehat{\mathfrak{C}}(%
\mathfrak{U}^{\ast }\cdot ,\mathfrak{U}^{\ast }\cdot )$, where $\mathfrak{U}%
^{\ast }$ is the adjoint of $\mathfrak{U}$ (i.e.~the unique operator $%
\mathfrak{U}^{\ast }$ that satisfies $\langle f,\mathfrak{U}g\rangle
=\langle \mathfrak{U}^{\ast }f,g\rangle $ for every $f$ and $g$ in $\mathcal{%
H}$.) The rotation equivariance of $\widehat{\mathfrak{C}}$ automatically
implies rotation equivariance of the principal component estimators obtained
from $\widehat{\mathfrak{C}}$.

Our trimmed estimators satisfy these properties, as shown in Proposition \ref%
{prop:Equivariance}. This is a consequence of the translation and rotation
invariance of the radii, and therefore of the weights $w(X_{i})$ (which, in
addition, are scale invariant). Note that translation, scale and rotation
invariance are properties that any \textquotedblleft
outlyingness\textquotedblright\ measure should satisfy: if an observation is
considered an outlier for a given dataset, the same observation should still
be considered an outlier if the dataset is simply translated, rotated or
re-scaled.

\begin{proposition}
\label{prop:Equivariance}Let $X_{1},\ldots ,X_{n}$ be a sample in $\mathcal{H%
}$, $a\neq 0$ a scalar, $b\in \mathcal{H}$, and $\mathfrak{U}$ a unitary
operator. Let $\tilde{X}_{i}=a\mathfrak{U}X_{i}+b$; denote by $\{\tilde{d}%
_{ij}\}$ and $\{\tilde{r}_{i}\}$ the corresponding interdistances and radii,
and by $\widehat{\tilde{\mu}}$, $\widehat{\mathfrak{\tilde{C}}}$, $\{%
\widehat{\tilde{\lambda}}_{k}\}$ and $\{\widehat{\tilde{\phi}}_{k}\}$\ the
corresponding estimators. Then:

\begin{enumerate}
\item $\tilde{d}_{ij}=|a|d_{ij}$ for all $i$ and $j$, and $\tilde{r}%
_{i}=\left\vert a\right\vert r_{i}$ for all $i$. Therefore $\mathrm{rank}(%
\tilde{r}_{i})=\mathrm{rank}(r_{i})$ and $w(\tilde{X}_{i})=w(X_{i})$ for all 
$i$.

\item $\widehat{\tilde{\mu}}=a\mathfrak{U}\hat{\mu}+b$.

\item $\widehat{\mathfrak{\tilde{C}}}(f,g)=a^{2}\widehat{\mathfrak{C}}(%
\mathfrak{U}^{\ast }f,\mathfrak{U}^{\ast }g)$ for all $f$ and $g$. Therefore 
$\widehat{\tilde{\lambda}}_{k}=a^{2}\hat{\lambda}_{k}$ and $\widehat{\tilde{%
\phi}}_{k}=\mathfrak{U}\hat{\phi}_{k}$ for all $k$ (note that the order of
the principal components is preserved).
\end{enumerate}
\end{proposition}

The robustness of an estimator is usually measured by the breakdown point
(Donoho and Huber 1983). The finite-sample breakdown point is the largest
proportion of outliers that an estimator can tolerate. More rigorously:
given a sample $\mathcal{X}=\{X_{1},\ldots ,X_{n}\}$, let $\mathcal{\tilde{X}%
}_{k}$ be a contaminated sample obtained from $\mathcal{X}$ by changing $k$
points arbitrarily; then the finite-sample breakdown point of $\hat{\mu}$ is 
$\varepsilon _{n}^{\ast }(\hat{\mu}):=k^{\ast }/n$, where $k^{\ast }$ is the
smallest $k$ for which there is a sequence of contaminated samples $\{%
\mathcal{\tilde{X}}_{k}^{(m)}\}_{m\geq 1}$ such that $\Vert \hat{\mu}%
^{(m)}\Vert \underset{m\rightarrow \infty }{\longrightarrow }\infty $. The
finite-sample breakdown point of $\widehat{\mathfrak{C}}$ is defined
analogously. The asymptotic breakdown point is the limit of $\varepsilon
_{n}^{\ast }(\hat{\mu})$ as $n$ goes to infinity, if the limit exists. The
highest asymptotic breakdown point attainable by an equivariant estimator is
.50 (Lopuha\"{a} and Rousseeuw 1991).

\begin{proposition}
\label{prop:BDP}Suppose $w(X_{i})=g(\mathrm{rank}(r_{i})/n)$, with $g$
satisfying the conditions given in Section \ref{sec:Definition}. If $\alpha
\leq .50$, $\lceil \alpha n\rceil \geq 3$, and $\beta \leq .50$, then $%
\varepsilon _{n}^{\ast }(\hat{\mu})=\varepsilon _{n}^{\ast }(\widehat{%
\mathfrak{C}})=\min (\lceil \alpha n\rceil ,\lfloor \beta n\rfloor +2)/n$,
which tends to $\min (\alpha ,\beta )$ when $n$ goes to infinity.
\end{proposition}

This proposition shows that the asymptotic breakdown point of the trimmed
estimators is $\min (\alpha ,\beta )$. Then, if $\alpha =.50$, the breakdown
point is just the trimming proportion $\beta $, and the optimal breakdown
point can be attained with $\beta =.50$. In practice, though, such
estimators are very inefficient when the actual proportion of outliers is
much less than 50\%, as we will show by simulation in Section \ref%
{sec:Simulations}. A better alternative is to use \textquotedblleft
soft\textquotedblright\ trimming, as explained in Section \ref%
{sec:Definition}.

\subsection{Population versions and properties}

The estimators (\ref{eq:mu_hat_gen}) and (\ref{eq:C_hat_gen}) can be
generalized to any probability measure $P$ on $\mathcal{H}$, of which (\ref%
{eq:mu_hat_gen}) and (\ref{eq:C_hat_gen}) can be seen as particular cases
obtained for $P=P_{n}$, the empirical measure on the sample $\{X_{1},\ldots
,X_{n}\}$. One of the reasons this generalization is useful is that it
allows us to study the consistency of the estimators: since $%
P_{n}\rightarrow P$ when the $X_{i}$s are i.i.d.~with distribution $P$,
under certain conditions (Fernholz 1983; Van der Vaart 1998, ch.~20) $\hat{%
\mu}$ and $\widehat{\mathfrak{C}}$ will converge in probability to their
respective population versions $\mu _{P}$ and $\mathfrak{C}_{P}$.

The derivation of $\mu _{P}$ and $\mathfrak{C}_{P}$ is as follows. Let $X$
be a stochastic process with distribution $P$. Define $F_{P}(t;v)=P\{\Vert
X-v\Vert \leq t\}$ for each $v\in \mathcal{H}$. The radius of the smallest
ball centered at $v$ with probability $\alpha $ is $r_{P}(v)=F_{P}^{-1}(%
\alpha ;v)$, where $F_{P}^{-1}(\alpha ;v):=\min \{t:F_{P}(t;v)\geq \alpha \}$
is the usual quantile function. Then $r_{P}(X)$ is the $\alpha $-radius
around $X$, and if $G_{P}(t):=P\{r_{P}(X)\leq t\}$, the weight function $%
w_{P}(v)$ has the form $w_{P}(v)=g[G_{P}\{r_{P}(v)\}]$, with $g$ as in
Section \ref{sec:Definition}. Then 
\begin{equation*}
\mu _{P}=\frac{E_{P}\{w_{P}(X)X\}}{E_{P}\{w_{P}(X)\}}
\end{equation*}%
and 
\begin{equation*}
\mathfrak{C}_{P}(f,g)=\frac{E_{P}\{w_{P}(X)\langle X-\mu _{P},f\rangle
\langle X-\mu _{P},g\rangle \}}{E_{P}\{w_{P}(X)\}}.
\end{equation*}%
The eigenvalues and eigenfunctions of $\mathfrak{C}_{P}$ will be denoted by $%
\lambda _{k,P}$ and $\phi _{k,P}$, respectively.

The following proposition shows that $\mu _{P}$ and $\mathfrak{C}_{P}$ are
well-defined for \emph{any} probability distribution $P$ on $\mathcal{H}$,
even if $\Vert X\Vert $ does not have finite moments of any order.

\begin{proposition}
\label{lem:Estim_well_defined}For any $\alpha >0$ there is a constant $%
K_{\alpha ,P}\geq 0$ such that $\Vert v\Vert \leq r_{P}(v)+K_{\alpha ,P}$
for all $v\in \mathcal{H}$. Therefore, if $\beta >0$ then $%
E_{P}\{w_{P}(X)\Vert X\Vert ^{k}\}<\infty $ for any $k\geq 0$.
\end{proposition}

The next proposition shows that $r_{P}(v)$ is really a measure of
outlyingness, in the sense that $r_{P}(v)$ is larger in regions of $\mathcal{%
H}$ where $P$ is less concentrated.

\begin{proposition}
\label{prop:r_depth}If $v$ and $w$ are two points in $\mathcal{H}$ such that 
$P(B_{\delta }(v))\geq P(B_{\delta }(w))$ for all $\delta >0$ (where $%
B_{\delta }(v)$ denotes the ball with center $v$ and radius $\delta $), then 
$r_{P}(v)\leq r_{P}(w)$.
\end{proposition}

The equivariance of $\hat{\mu}$ and $\widehat{\mathfrak{C}}$ carries over to 
$\mu _{P}$ and $\mathfrak{C}_{P}$ (the proof is given in the technical
supplement). A consequence of the translation equivariance of $\mu _{P}$ is
the following:

\begin{proposition}
\label{prop:Mu_P_symmetric_dist}If $X$ has a symmetric distribution about $%
\mu _{0}$ (i.e.~if $X-\mu _{0}$ and $\mu _{0}-X$ are identically
distributed), then $\mu _{P}=\mu _{0}$.
\end{proposition}

To study the population versions of the trimmed principal components let us
assume that $X$ admits, with probability 1, the decomposition 
\begin{equation}
X=\mu _{0}+\sum_{k\in \mathcal{I}}\lambda _{0k}^{1/2}Z_{k}\phi _{0k},
\label{eq:K-L_decomposition}
\end{equation}%
where $\mu _{0}\in \mathcal{H}$, the $Z_{k}$s are real random variables, $%
\{\phi _{0k}\}\subset \mathcal{H}$ is an orthonormal system, and $\{\lambda
_{0k}\}$ is a strictly positive non-increasing sequence with $\sum_{k\in 
\mathcal{I}}\lambda _{0k}<\infty $; the set of indices $\mathcal{I}$ is
countable but may be finite or infinite. This decomposition holds, for
instance, if $E(\Vert X\Vert ^{2})<\infty $, and is known as the Karhunen--Lo%
\`{e}ve decomposition (Ash and Gardner 1975, ch.~1.4). In that case $%
E(X)=\mu _{0}$, the $\phi _{0k}$s and the $\lambda _{0k}$s are the
eigenfunctions and eigenvalues of the covariance operator, and $%
Z_{k}=\langle X-\mu _{0},\phi _{0k}\rangle /\lambda _{0k}^{1/2}$ are
uncorrelated with $E(Z_{k})=0$ and $\mathrm{var}(Z_{k})=1$.

But expansion (\ref{eq:K-L_decomposition}) also holds in some situations
where $E(\Vert X\Vert ^{2})=\infty $, providing a meaningful notion of
\textquotedblleft heavy-tailed distributions\textquotedblright\ for
functional spaces. For instance, if the $Z_{k}$s in (\ref%
{eq:K-L_decomposition}) are independent, Kolmogorov's Three Series Theorem
(Gikhman and Skorokhod 2004, p.~384) implies that $\sum_{k\in \mathcal{I}%
}\lambda _{0k}^{1/2}Z_{k}\phi _{0k}$ converges almost surely in $\mathcal{H}$
if and only if $\sum_{k\in \mathcal{I}}P(\lambda _{0k}Z_{0k}^{2}>c)<\infty $
for every $c>0$. The latter is satisfied whenever the $\lambda _{0k}$s go to
zero fast enough, even if the $Z_{k}$s do not have finite moments of any
order. For example, if the $Z_{k}$s have a Cauchy distribution, 
\begin{equation*}
\sum_{k\in \mathcal{I}}P(\lambda _{0k}Z_{k}^{2}>c)\leq \sum_{k\in \mathcal{I}%
}\frac{2}{\pi }\left( \frac{\lambda _{0k}}{c}\right) ^{1/2},
\end{equation*}%
and the right-hand side is finite for any $c>0$ as long as $\sum_{k\in 
\mathcal{I}}\lambda _{0k}^{1/2}<\infty $.

Model (\ref{eq:K-L_decomposition}) is also useful to characterize the two
types of outliers that may be present in a functional sample. One type of
outliers would be observations that satisfy model (\ref{eq:K-L_decomposition}%
) but with extreme values of the $Z_{k}$s, which we call \emph{intrinsic}
outliers, since they belong to the space generated by the $\phi _{0k}$s.
Another type of outliers would be those that do not follow model (\ref%
{eq:K-L_decomposition}) at all, which we denominate \emph{extrinsic}
outliers, since they fall outside the subspace of $\mathcal{H}$ where the
\textquotedblleft good\textquotedblright\ data lives. To exemplify the
difference: suppose that a sample of curves shows a prominent feature, such
as a peak, and the leading principal component $\phi _{01}$ explains
variability around this peak (a usual situation). An intrinsic outlier would
be a curve with an unusual peak (either too flat or too sharp compared to
the other curves), whereas an extrinsic outlier would be an observation with
a peak at a different location, where the rest of the data shows no such
feature. Our estimators can handle both types of outliers, since the
interdistances make no distinction between the two types (although extrinsic
outliers are easier to spot). The outliers considered in the simulations
(Section \ref{sec:Simulations}) are intrinsic outliers, while those in the
examples (Section \ref{sec:Examples}) are mostly extrinsic outliers.

Note that under model (\ref{eq:K-L_decomposition}) the interdistances
satisfy $d_{ij}^{2}=\sum_{k\in \mathcal{I}}\lambda _{0k}(Z_{ki}-Z_{kj})^{2}$%
, so the distribution of the $d_{ij}$s (and therefore of the radii) depends
entirely on the $Z_{k}$s and the $\lambda _{0k}$s, not on $\mu _{0}$ or the $%
\phi _{0k}$s. This implies the following:

\begin{proposition}
\label{prop:C(P)_spectral_decomp}If expansion (\ref{eq:K-L_decomposition})
holds with independent and symmetrically distributed $Z_{k}$s, then 
\begin{equation}
\mathfrak{C}_{P}(f,g)=\sum_{k\in \mathcal{I}}\tilde{\lambda}_{0k}\langle
\phi _{0k},f\rangle \langle \phi _{0k},g\rangle  \label{eq:C_P_expansion}
\end{equation}%
with 
\begin{equation}
\tilde{\lambda}_{0k}=\frac{E_{P}\{w_{P}(X)|\langle X-\mu _{0},\phi
_{0k}\rangle |^{2}\}}{E_{P}\{w_{P}(X)\}}=\lambda _{0k}\ \frac{%
E_{P}\{w_{P}(X)Z_{k}^{2}\}}{E_{P}\{w_{P}(X)\}}.  \label{eq:lambda_tilde}
\end{equation}
The sequence $\{\tilde{\lambda}_{0k}\}$ is strictly positive but not
necessarily decreasing, and it depends entirely on the distribution of $%
\{\lambda _{0k}^{1/2}Z_{k}\}$. In addition, if the $Z_{k}$s are identically
distributed, then $\lambda _{0j}=\lambda _{0k}$ implies $\tilde{\lambda}%
_{0j}=\tilde{\lambda}_{0k}$, so that the multiplicity of the eigenvalues is
preserved.
\end{proposition}

This result implies that the set of principal components of $\mathfrak{C}%
_{P} $, $\{\phi _{k,P}\}$, coincides with the set $\{\phi _{0k}\}$, but it
cannot be said in general that $\phi _{k,P}=\phi _{0k}$ for each $k$,
because the sequence $\{\tilde{\lambda}_{0k}\}$ is not necessarily
decreasing. The reason is that although $\lambda _{0k}Z_{k}^{2}$ is
stochastically greater than $\lambda _{0j}Z_{j}^{2}$ when $\lambda
_{0k}>\lambda _{0j}$ and the $Z_{k}$s are identically distributed, this does
not imply that $w_{P}(X)\lambda _{0k}Z_{k}^{2}$ is stochastically greater
than $w_{P}(X)\lambda _{0j}Z_{j}^{2}$ in general, so it cannot be guaranteed
that $\tilde{\lambda}_{0k}\geq \tilde{\lambda}_{0j}$. However, (\ref%
{eq:lambda_tilde}) does imply that $\tilde{\lambda}_{0k}>0$ if and only if $%
\lambda _{0k}>0$, so the dimension of the model is preserved.

\section{Simulations\label{sec:Simulations}}

We ran a Monte Carlo study to assess the comparative performance of the
following estimators:

\begin{itemize}
\item The sample mean and sample principal components.

\item The spatial median and spherical principal components (Locantore et
al.~1999, Gervini 2008). The spatial median is defined as the $\hat{\mu}$
that minimizes $\sum_{i=1}^{n}\Vert X_{i}-\mu \Vert $, and the spherical
principal components are defined as the principal components of the
normalized sample $\{\left( X_{i}-\hat{\mu}\right) /\left\Vert X_{i}-\hat{\mu%
}\right\Vert \}$, i.e.~the eigenfunctions of the covariance operator 
\begin{equation*}
\widehat{\mathfrak{C}}(f,g):=\frac{1}{n}\sum_{i=1}^{n}\langle \frac{X_{i}-%
\hat{\mu}}{\left\Vert X_{i}-\hat{\mu}\right\Vert },f\rangle \langle \frac{%
X_{i}-\hat{\mu}}{\left\Vert X_{i}-\hat{\mu}\right\Vert },g\rangle .
\end{equation*}

\item Trimmed estimators based on the deviations $\left\Vert X_{i}-\hat{\mu}%
\right\Vert $, where $\hat{\mu}$ is the spatial median, with 20\% and 50\%
trimming. The observations with the largest 20\% or 50\% deviations where
eliminated and the mean and principal components of the remaining data was
computed.

\item Trimmed estimators based on $h$-depth (Cuevas et al.~2007), with 20\%
and 50\% trimming. The $h$-depth of a datum $z$ is defined as 
\begin{equation*}
\hat{f}_{h}(z)=\frac{1}{n}\sum_{i=1}^{n}K_{h}(\Vert z-X_{i}\Vert ),
\end{equation*}%
where $K_{h}(t)=h^{-1}K(t/h)$ for some kernel function $K$. Following Cuevas
et al.~(2007), we take $K$ as the Gaussian density and $h$ as the 20th
percentile of the set of $L_{2}$-interdistances (there is no clear rationale
for this choice but we used the same tuning parameters as Cuevas et al.~in
order to make our simulation results comparable to theirs). Note that a
small, not a large, value of $\hat{f}_{h}(X_{i})$ would indicate that $X_{i}$
is an outlier, so we trim those observations with small value of $\hat{f}%
_{h} $. In the extensive simulations run by Cuevas et al., these estimators
outperformed the estimators of Fraiman and Muniz (2001) and some
projection-based estimators, so we did not include the latter in our
simulations.

\item Trimmed estimators based on band depth (L\'{o}pez-Pintado and Romo
2009), with 20\% and 50\% trimming. The band depth is computed as follows.
Given real-valued functions $f_{1},\ldots ,f_{k}$ defined on some interval $%
I\subseteq \mathbb{R}$, with $k\geq 2$, the $k$-band\ spanned by these
functions is 
\begin{equation*}
B(f_{1},\ldots ,f_{k}):=\{(t,y):t\in I,y\in \lbrack \min_{1\leq i\leq
k}f_{i}(t),\max_{1\leq i\leq k}f_{i}(t)]\}.
\end{equation*}%
For a given curve $z$ and a sample $\mathcal{X}=\{X_{1},\ldots ,X_{n}\}$,
let $BD_{k}(z;\mathcal{X)}$ be the average number of sample $k$-bands that
contain the graph of $z$; that is, 
\begin{equation*}
BD_{k}(z;\mathcal{X)=}\binom{n}{k}^{-1}\sum_{1\leq i_{1}<\cdots <i_{k}\leq n}%
\mathbb{I}\{G(z)\subseteq B(X_{i_{1}},\ldots ,X_{i_{k}})\},
\end{equation*}%
where $G(z):=\{(t,z(t)):t\in I\}$. The $J$-depth of the curve $z$ is defined
as $D_{J}(z;\mathcal{X})=\sum_{k=2}^{J}BD_{k}(z;\mathcal{X)}$. As
recommended by L\'{o}pez-Pintado and Romo (2009), we use $J=3$. Once again,
outliers are indicated by small values of $D_{J}(z;\mathcal{X})$, so we trim
the $100\beta \%$ observations with smallest values of $D_{J}$.

\item The trimmed estimators introduced in this article, with hard and soft
rejection weights. For hard-rejection weights, 20\% and 50\% fixed trimming
was considered as well as the adaptive $\beta $ estimated with Gamma
mixtures; for the soft-rejection weight (\ref{eq:weight_function_g}), the
parameters $\beta _{1}=.50$ and $\beta =.20$ were used. In all cases, the
radii were computed with $\alpha =.50$ (simulations with $\alpha =.20$ were
also run but not reported here, because the estimator's performance was
uniformly worse than for $\alpha =.50$).
\end{itemize}

The data was generated following model (\ref{eq:K-L_decomposition}) with $%
\mu _{0}=0$ and $\phi _{0k}(t)=\sqrt{2}\sin (\pi kt)$, for $t\in \lbrack
0,1] $. The $Z_{k}$s followed different distributions for each scenario, as
explained below. Two sequences of eigenvalues were considered: a
slow-decaying sequence $\lambda _{0k}=1/\{k(k+1)\}$ (Model 1), and a
fast-decaying sequence $\lambda _{0k}=1/2^{k}$ (Model 2); note that $%
\sum_{k=1}^{\infty }\lambda _{k}=1$ in both cases. Model 2 is practically a
finite-dimensional model, since the first five terms accumulate 97\% of the
variability; Model 1, on the other hand, needs 31 terms to accumulate the
same proportion of the variability, so it is an infinite-dimensional\ model
for practical purposes. For actual data generation we truncated Model 1 at
the 1000th term and Model 2 at the 10th term, which represent 99.9\% of the
total variability in both cases. The sample size was $n=50$ in all cases,
and the curves were discretized at an equally spaced grid of 100 points.
Each sampling situation was replicated 2000 times; the mean absolute errors
reported in Tables \ref{tab:Simulations_Mean_K3} and \ref%
{tab:Simulations_PC_K3} are accurate up to two significant places (we do not
report Monte Carlo standard errors for reasons of space and readability).

Regarding the distribution of the $Z_{k}$s, we were interested in three
situations: \emph{(i)} non-contaminated Normal data, \emph{(ii)}
outlier-contaminated Normal data, and \emph{(iii)} non-Normal data
(specifically, data with heavier tails than Normal). For case \emph{(i)} we
generated i.i.d.$~Z_{k}$s with $N(0,1)$ distribution. For case \emph{(ii)}
we considered two scenarios: to study the robustness of the location
estimators, we generated outliers by adding $3\phi _{01}(t)$ to $%
n\varepsilon $ sample curves (which creates a bias in $\hat{\mu}$); to study
the robustness of the estimators of $\phi _{1}$, we generated outliers by
adding $3\phi _{02}(t)$ to $n\varepsilon /2$ sample curves and subtracting
the same quantity from other $n\varepsilon /2$ sample curves (this
contamination inflates the variability in the direction of $\phi _{02}$,
creating a bias in $\hat{\phi}_{1}$ without affecting $\hat{\mu}$). Four
values of $\varepsilon $ were considered: $.10$, $.20$, $.30$, and $.40$.
For case \emph{(iii)} we generated i.i.d.~$Z_{k}$s with Student's $t_{\nu }$
distribution, with degrees of freedom $\nu $ equal to 1, 2, and 3.

%TCIMACRO{%
%\TeXButton{B}{\begin{table}
%\centering}}%
%BeginExpansion
\begin{table}
\centering%
%EndExpansion
%TCIMACRO{\TeXButton{TeX field}{\begin{sideways}}}%
%BeginExpansion
\begin{sideways}%
%EndExpansion
\begin{tabular}{ccccccccccccc}
\hline
&  & Normal & \multicolumn{4}{c}{Contaminated Normal} &  & 
\multicolumn{3}{c}{Student} &  & Ranking \\ 
\cline{4-7}\cline{4-7}\cline{9-11}
Model & Estimator &  & 10\% & 20\% & 30\% & 40\% &  & $t_{1}$ & $t_{2}$ & $%
t_{3}$ &  &  \\ \hline
1 & \multicolumn{1}{l}{Mean} & .134 & .318 & .607 & .906 & 1.206 &  & 74.44
& .406 & .225 &  & 9.8 \\ 
& \multicolumn{1}{l}{Median} & .140 & .187 & .320 & .539 & .886 &  & 1.00 & 
.239 & .190 &  & 5.6 \\ 
& \multicolumn{1}{l}{Deviation (20\%)} & .158 & .165 & .172 & .343 & .739 & 
& 1.51 & .260 & .204 &  & 5.4 \\ 
& \multicolumn{1}{l}{Deviation (50\%)} & .181 & .209 & .262 & .325 & .498 & 
& 1.05 & .275 & .229 &  & 7.2 \\ 
& \multicolumn{1}{l}{$h$-depth (20\%)} & .159 & .160 & .171 & .404 & .867 & 
& 1.54 & .257 & .205 &  & 5.8 \\ 
& \multicolumn{1}{l}{$h$-depth (50\%)} & .185 & .190 & .201 & .226 & .373 & 
& 1.06 & .274 & .229 &  & 6.2 \\ 
& \multicolumn{1}{l}{Band depth (20\%)} & .150 & .152 & .169 & .370 & .725 & 
& 34.66 & .451 & .252 &  & 6.9 \\ 
& \multicolumn{1}{l}{Band depth (50\%)} & .189 & .214 & .268 & .347 & .454 & 
& 23.63 & .518 & .318 &  & 9.8 \\ 
& \multicolumn{1}{l}{Hard trimmed (20\%)} & .165 & .164 & .166 & .288 & .634
&  & 1.45 & .256 & .208 &  & 4.5 \\ 
& \multicolumn{1}{l}{Hard trimmed (50\%)} & .197 & .198 & .201 & .210 & .283
&  & 1.06 & .281 & .238 &  & 6.6 \\ 
& \multicolumn{1}{l}{Hard trimmed (adap.)} & .169 & .157 & .180 & .220 & .410
&  & 54.04 & .360 & .223 &  & 6.0 \\ 
& \multicolumn{1}{l}{Soft trimmed (20\%)} & .175 & .175 & .177 & .198 & .396
&  & 1.10 & .253 & .211 &  & 4.4 \\ 
&  &  &  &  &  &  &  &  &  &  &  &  \\ 
2 & \multicolumn{1}{l}{Mean} & .132 & .322 & .606 & .906 & 1.20 &  & 9.423 & 
.384 & .223 &  & 10.1 \\ 
& \multicolumn{1}{l}{Median} & .141 & .192 & .317 & .530 & .871 &  & .348 & 
.208 & .180 &  & 4.6 \\ 
& \multicolumn{1}{l}{Deviation (20\%)} & .161 & .164 & .175 & .352 & .728 & 
& .506 & .224 & .194 &  & 4.6 \\ 
& \multicolumn{1}{l}{Deviation (50\%)} & .183 & .207 & .258 & .318 & .488 & 
& .369 & .235 & .210 &  & 6.6 \\ 
& \multicolumn{1}{l}{$h$-depth (20\%)} & .162 & .162 & .174 & .434 & .908 & 
& .547 & .228 & .194 &  & 5.2 \\ 
& \multicolumn{1}{l}{$h$-depth (50\%)} & .185 & .193 & .204 & .220 & .427 & 
& .369 & .235 & .215 &  & 5.9 \\ 
& \multicolumn{1}{l}{Band depth (20\%)} & .149 & .243 & .438 & .695 & .985 & 
& 1.112 & .281 & .222 &  & 8.9 \\ 
& \multicolumn{1}{l}{Band depth (50\%)} & .173 & .416 & .630 & .877 & 1.18 & 
& 1.166 & .363 & .278 &  & 10.9 \\ 
& \multicolumn{1}{l}{Hard trimmed (20\%)} & .166 & .166 & .167 & .292 & .640
&  & .478 & .226 & .195 &  & 4.5 \\ 
& \multicolumn{1}{l}{Hard trimmed (50\%)} & .195 & .197 & .203 & .208 & .304
&  & .377 & .240 & .221 &  & 6.3 \\ 
& \multicolumn{1}{l}{Hard trimmed (adap.)} & .170 & .158 & .182 & .229 & .429
&  & 10.325 & .336 & .216 &  & 6.4 \\ 
& \multicolumn{1}{l}{Soft trimmed (20\%)} & .175 & .177 & .179 & .199 & .413
&  & .377 & .221 & .200 &  & 4.1 \\ \hline
\end{tabular}%
%TCIMACRO{\TeXButton{TeX field}{\end{sideways}}}%
%BeginExpansion
\end{sideways}%
%EndExpansion
\caption{Simulation Results. Mean absolute errors of location estimators.}%
\label{tab:Simulations_Mean_K3}%
%TCIMACRO{\TeXButton{E}{\end{table}}}%
%BeginExpansion
\end{table}%
%EndExpansion

Table \ref{tab:Simulations_Mean_K3} reports the mean absolute error $E(\Vert 
\hat{\mu}-\mu _{0}\Vert )$ for each estimator and each sampling
distribution. Since there are 12 estimators and 8 sampling distributions it
is hard to make conclusions at a glance. To facilitate comparisons, we
ranked the estimators in increasing order of error for each sampling
distribution, and computed the average rank for each estimator; this average
rank is given in the last column of Table \ref{tab:Simulations_Mean_K3}. We
see that the comparative performance of the estimators is similar under both
models. The soft-trimmed estimators show the best overall performance, since
they have the smallest average ranks; in the other extreme, the band-depth
50\%-trimmed estimators show the worst overall performance. Looking into the
numbers in more detail, we see that hard-trimmed estimators with 20\%
trimming perform poorly for contaminated Normal distributions with $%
\varepsilon >.20$ and for the Cauchy distribution. Among hard-trimmed
estimators with 50\% trimming, our estimator and the $h$-depth-based
estimator are comparable, the former being better for contaminated Normals
with $\varepsilon \geq .20$ (and significantly better for $\varepsilon =.40$%
) and the latter being slightly better in the other situations. The
soft-trimmed estimator shows an intermediate behavior between the 20\% and
50\% hard-trimmed estimators; even at the most extreme cases of the 40\%
contaminated Normal and the Cauchy distribution, its estimation error is not
much larger than that of the 50\% hard-trimmed estimator. The adaptive
estimator also shows an intermediate behavior between the 20\% and 50\%
hard-trimmed estimators, except for the Cauchy distribution, for which it
does not even seem to be well defined (the same can be said for the
estimators based on band depth); this is not entirely surprising, since the
Cauchy distribution produces a single heavy-tailed distribution of radii
rather than a mixture. All things considered, the soft-trimmed estimator
seems to offer the best trade-off between robustness and efficiency.

For the principal component estimators, the mean absolute errors $E(\Vert 
\hat{\phi}_{1}-\phi _{01}\Vert )$ are reported in Table \ref%
{tab:Simulations_PC_K3}. Breakdown of $\hat{\phi}_{1}$ occurs when $\hat{\phi%
}_{1}$ is orthogonal to $\phi _{01}$, in which case $\Vert \hat{\phi}%
_{1}-\phi _{01}\Vert =\sqrt{2}$, so the errors are always bounded. The
best-ranked estimators are now the spherical principal components, which is
rather unexpected, but looking into the numbers in more detail, we see that
this is mostly because of their low errors for $t$ distributions. Their
performance for contaminated Normal distributions is not good, showing very
large errors for contamination proportions as small as 20\%. Our
hard-trimmed estimators and the $h$-depth-based estimators show comparable
performances, although once again the soft-trimmed estimator offers a better
trade-off between robustness and efficiency: although it breaks down for the
40\%-contaminated Normal, it has much lower estimation errors than the
50\%-hard-trimmed estimators for lower levels of contamination and for $t$
distributions (even for the Cauchy). The adaptive estimator does no break
down for the 40\%-contaminated Normal, but it does for the Cauchy
distribution.

%TCIMACRO{\TeXButton{B}{\begin{table}[tbp] \centering}}%
%BeginExpansion
\begin{table}[tbp] \centering%
%EndExpansion
%TCIMACRO{\TeXButton{TeX field}{\begin{sideways}}}%
%BeginExpansion
\begin{sideways}%
%EndExpansion
\begin{tabular}{ccccccccccccc}
\hline
&  & Normal & \multicolumn{4}{c}{Contaminated Normal} &  & 
\multicolumn{3}{c}{Student} &  & Ranking \\ 
\cline{4-7}\cline{4-7}\cline{9-11}
Model & Estimator &  & 10\% & 20\% & 30\% & 40\% &  & $t_{1}$ & $t_{2}$ & $%
t_{3}$ &  &  \\ \hline
1 & \multicolumn{1}{l}{Sample p.c.} & .168 & 1.27 & 1.36 & 1.37 & 1.38 &  & 
1.25 & .508 & .263 &  & 8.5 \\ 
& \multicolumn{1}{l}{Spherical p.c.} & .204 & .297 & .879 & 1.24 & 1.34 &  & 
.577 & .268 & .232 &  & 4.8 \\ 
& \multicolumn{1}{l}{Deviation (20\%)} & .322 & .280 & .203 & 1.25 & 1.36 & 
& .998 & .393 & .350 &  & 6.0 \\ 
& \multicolumn{1}{l}{Deviation (50\%)} & .644 & .615 & .543 & .492 & .390 & 
& .912 & .598 & .604 &  & 8.5 \\ 
& \multicolumn{1}{l}{$h$-depth (20\%)} & .300 & .271 & .210 & 1.27 & 1.36 & 
& .972 & .373 & .327 &  & 5.3 \\ 
& \multicolumn{1}{l}{$h$-depth (50\%)} & .526 & .511 & .492 & .481 & .414 & 
& .898 & .545 & .536 &  & 7.5 \\ 
& \multicolumn{1}{l}{Band depth (20\%)} & .192 & .199 & .222 & 1.28 & 1.35 & 
& 1.26 & .541 & .304 &  & 6.1 \\ 
& \multicolumn{1}{l}{Band depth (50\%)} & .256 & .326 & .455 & .552 & .738 & 
& 1.24 & .604 & .396 &  & 7.5 \\ 
& \multicolumn{1}{l}{Hard trimmed (20\%)} & .310 & .288 & .218 & 1.12 & 1.36
&  & .979 & .400 & .347 &  & 5.9 \\ 
& \multicolumn{1}{l}{Hard trimmed (50\%)} & .467 & .478 & .473 & .479 & .441
&  & .902 & .560 & .532 &  & 7.3 \\ 
& \multicolumn{1}{l}{Hard trimmed (adap.)} & .331 & .224 & .247 & .271 & .344
&  & 1.20 & .520 & .361 &  & 5.1 \\ 
& \multicolumn{1}{l}{Soft trimmed (20\%)} & .347 & .335 & .282 & .268 & 1.25
&  & .778 & .418 & .377 &  & 5.6 \\ 
&  &  &  &  &  &  &  &  &  &  &  &  \\ 
2 & \multicolumn{1}{l}{Sample p.c.} & .224 & 1.29 & 1.36 & 1.37 & 1.38 &  & 
.966 & .583 & .396 &  & 9.0 \\ 
& \multicolumn{1}{l}{Spherical p.c.} & .281 & .493 & 1.06 & 1.27 & 1.34 &  & 
.479 & .362 & .327 &  & 4.9 \\ 
& \multicolumn{1}{l}{Deviation (20\%)} & .439 & .387 & .287 & 1.25 & 1.36 & 
& .688 & .473 & .448 &  & 5.9 \\ 
& \multicolumn{1}{l}{Deviation (50\%)} & .697 & .673 & .636 & .575 & .490 & 
& .698 & .652 & .657 &  & 8.5 \\ 
& \multicolumn{1}{l}{$h$-depth (20\%)} & .392 & .366 & .310 & 1.29 & 1.36 & 
& .659 & .449 & .416 &  & 4.9 \\ 
& \multicolumn{1}{l}{$h$-depth (50\%)} & .569 & .577 & .572 & .574 & .564 & 
& .675 & .613 & .592 &  & 7.3 \\ 
& \multicolumn{1}{l}{Band depth (20\%)} & .311 & .485 & 1.05 & 1.32 & 1.36 & 
& .733 & .471 & .409 &  & 7.1 \\ 
& \multicolumn{1}{l}{Band depth (50\%)} & .491 & .947 & 1.02 & 1.06 & 1.21 & 
& .594 & .462 & .461 &  & 6.5 \\ 
& \multicolumn{1}{l}{Hard trimmed (20\%)} & .400 & .380 & .303 & 1.15 & 1.35
&  & .693 & .478 & .448 &  & 5.4 \\ 
& \multicolumn{1}{l}{Hard trimmed (50\%)} & .524 & .545 & .541 & .566 & .633
&  & .700 & .621 & .598 &  & 7.8 \\ 
& \multicolumn{1}{l}{Hard trimmed (adap.)} & .432 & .307 & .363 & .415 & .570
&  & .893 & .548 & .480 &  & 5.6 \\ 
& \multicolumn{1}{l}{Soft trimmed (20\%)} & .421 & .424 & .377 & .410 & 1.26
&  & .599 & .484 & .470 &  & 5.1 \\ \hline
\end{tabular}%
%TCIMACRO{\TeXButton{TeX field}{\end{sideways}}}%
%BeginExpansion
\end{sideways}%
%EndExpansion
\caption{Simulation Results. Mean absolute errors of first principal
component estimators.}\label{tab:Simulations_PC_K3}%
%TCIMACRO{\TeXButton{E}{\end{table}}}%
%BeginExpansion
\end{table}%
%EndExpansion

The similar behavior of the estimators based on the radii and those based on
the $h$-depth is not accidental, because both are based on metric notions of
data depth: the $\alpha $-radius measures the distance between a given datum
and its closest $\lceil \alpha n\rceil $th neighbor, while the $h$-depth
essentially counts the number of observations within a fixed distance of a
given datum; so these measures are, in a way, the dual of one another.
However, the $\alpha $-radii have certain advantages over the $h$-depth: the
parameter $\alpha $ that defines the radii is an interpretable quantity,
while the parameter $h$\ that defines the $h$-depth is an arbitrary
bandwidth with an unknown effect on the estimator's properties. Also, the
breakdown point of the estimators based on $\alpha $-radii is known, while
the breakdown point of the estimators based on $h$-depth is unknown.

In contrast with these metric notions of data depth, the band depth of L\'{o}%
pez-Pintado and Romo (2009) is not based on distances but on the number of
\textquotedblleft bands\textquotedblright\ that cover each sample function.
Therefore the trimmed estimators based on band depth behave very differently
(in fact, much worse) than those based on $\alpha $-radii or $h$-depth. Two
additional disadvantages of band-depth trimming are that the determination
of all the \textquotedblleft bands\textquotedblright\ that cover a given
curve is a combinatorial problem, which is unfeasible for large sample
sizes, and that generalizing the concept of \textquotedblleft
band\textquotedblright\ to Euclidean spaces beyond univariate curves is not
obvious.

\section{\label{sec:Examples}Examples}

\subsection{Excitation--Emission Matrices}

As explained in Mortensen and Bro (2006), enzyme cultivation processes often
require quick on-line adjustments that demand fast and reliable quality
control tools. Samples of the cultivation broth are typically taken at
regular time intervals, and enzyme activity is measured. The traditional
off-line chemical analyses determine enzyme activity directly and
accurately, but it may take hours or days to get the results back from the
laboratory. An alternative is to employ multi-channel fluorescence sensors
that produce immediate results in the form of excitation-emission matrices
(EEMs), although enzyme activity can be determined only indirectly from the
EEMs (via principal component regression or partial least squares).

Mortensen and Bro (2006) analyze a dataset of 338 EEMs, available at \newline
\textsf{http://www.models.life.ku.dk/research/data/}. A movie showing these
338 EEMs in quick succession is available on the author's website. A few
atypical EEMs can be spotted at the end of the movie. Taking logarithms of
the EEMs ameliorates the effect of the outliers to some extent, but not
completely, as Figure \ref{fig:EEM_samples} shows (a movie showing the
log-EEMs is also available on the author's website).

In principle, an EEM is a two-dimensional array consisting of light
intensity measured at certain excitation and emission wavelengths. Mortensen
and Bro (2006) use 15 excitation filters ranging from 270 to 550 nm, and 15
emission filters ranging from 310 to 590 nm; all filters have a maximum
half-width of 20 nm. Since emission wavelength must be longer than
excitation wavelength, the EEMs are actually triangular arrays: of the $%
15\times 15$ possible excitation/emission combinations, only 120 yield
actual measurements. This problem could be approached as a classical
multivariate problem of dimension $p=120$ and sample size $n=338$, and some
of the robust methods reviewed by Filzmoser et al.~(2009) for the
\textquotedblleft large $p$, small $n$\textquotedblright\ problem could be
applied. However, since light intensity is a continuous function of the
excitation and emission wavelengths, it is more appropriate and
statistically efficient to approach this problem as a functional-data
problem; the 120 measurements are just an arbitrary discretization of the
continuous surfaces $X(s,t)$, which live in $\mathcal{L}^{2}(\mathbb{R}^{2})$%
. This is a Euclidean space with inner product $\langle f,g\rangle =\tiint
f(s,t)g(s,t)\ ds\ dt$, the mean of $X$ is the bivariate function $\mu
(s,t)=E\{X(s,t)\}$ and the covariance operator of $X$ can be represented as 
\begin{equation*}
\mathfrak{C}(f,g)=\diiiint \rho (s_{1},t_{1},s_{2},t_{2})\ f(s_{1},t_{1})\
g(s_{2},t_{2})\ ds_{1}\ dt_{1}\ ds_{2}\ dt_{2},
\end{equation*}%
where $\rho (s_{1},t_{1},s_{2},t_{2})=\mathrm{cov}%
\{X(s_{1},t_{1}),X(s_{2},t_{2})\}$.

\FRAME{ftbpFU}{3.2301in}{2.4327in}{0pt}{\Qcb{Excitation--Emission Matrices.
Histogram of the radii with $\protect\alpha =.50$.}}{\Qlb{fig:EEM_hist_r}}{%
eem_hist_r.eps}{\special{language "Scientific Word";type
"GRAPHIC";maintain-aspect-ratio TRUE;display "USEDEF";valid_file "F";width
3.2301in;height 2.4327in;depth 0pt;original-width 5.8219in;original-height
4.3708in;cropleft "0";croptop "1";cropright "1";cropbottom "0";filename
'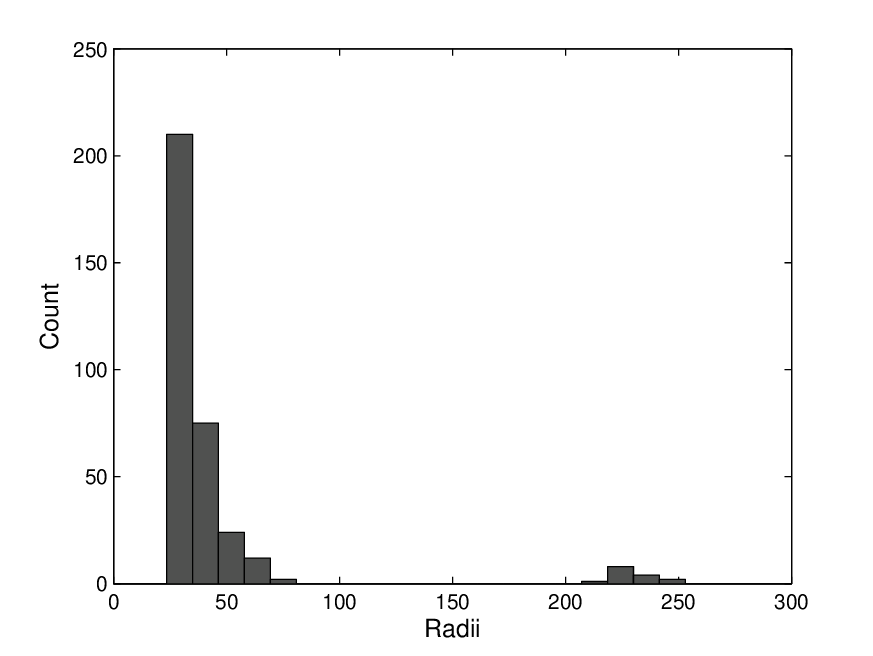';file-properties "XNPEU";}}

We carried out a principal component analysis on the log-EEMs. A histogram
of the radii (Figure \ref{fig:EEM_hist_r}) shows that 15 observations are
clear outliers, so we computed the 5\% trimmed mean and principal components
(Figure \ref{fig:EEM_mean_pcs}). Among the 20 leading components, the first
one (Figure \ref{fig:EEM_mean_pcs}(b)) accounts for 59\% of the variability,
and the second one (Figure \ref{fig:EEM_mean_pcs}(c)) accounts for 20\% of
the variability. We also computed the sample mean and principal components;
among the 20 leading components, the first one (Figure \ref{fig:EEM_mean_pcs}%
(e)) accounts for 88\% of the variability, and the second one (Figure \ref%
{fig:EEM_mean_pcs}(f)) for only 5\%.

\FRAME{ftbpFU}{6.1627in}{3.0614in}{0pt}{\Qcb{Excitation--Emission Matrices.
(a) Trimmed mean, (b) first trimmed principal component, (c) second trimmed
principal component, (d) sample mean, (e) first sample principal component,
and (f) second sample principal component. Trimmed estimators were computed
with 5\% trimming.}}{\Qlb{fig:EEM_mean_pcs}}{eem_mean_and_pcs.eps}{\special%
{language "Scientific Word";type "GRAPHIC";maintain-aspect-ratio
TRUE;display "USEDEF";valid_file "F";width 6.1627in;height 3.0614in;depth
0pt;original-width 10.4184in;original-height 5.1526in;cropleft "0";croptop
"1";cropright "1";cropbottom "0";filename
'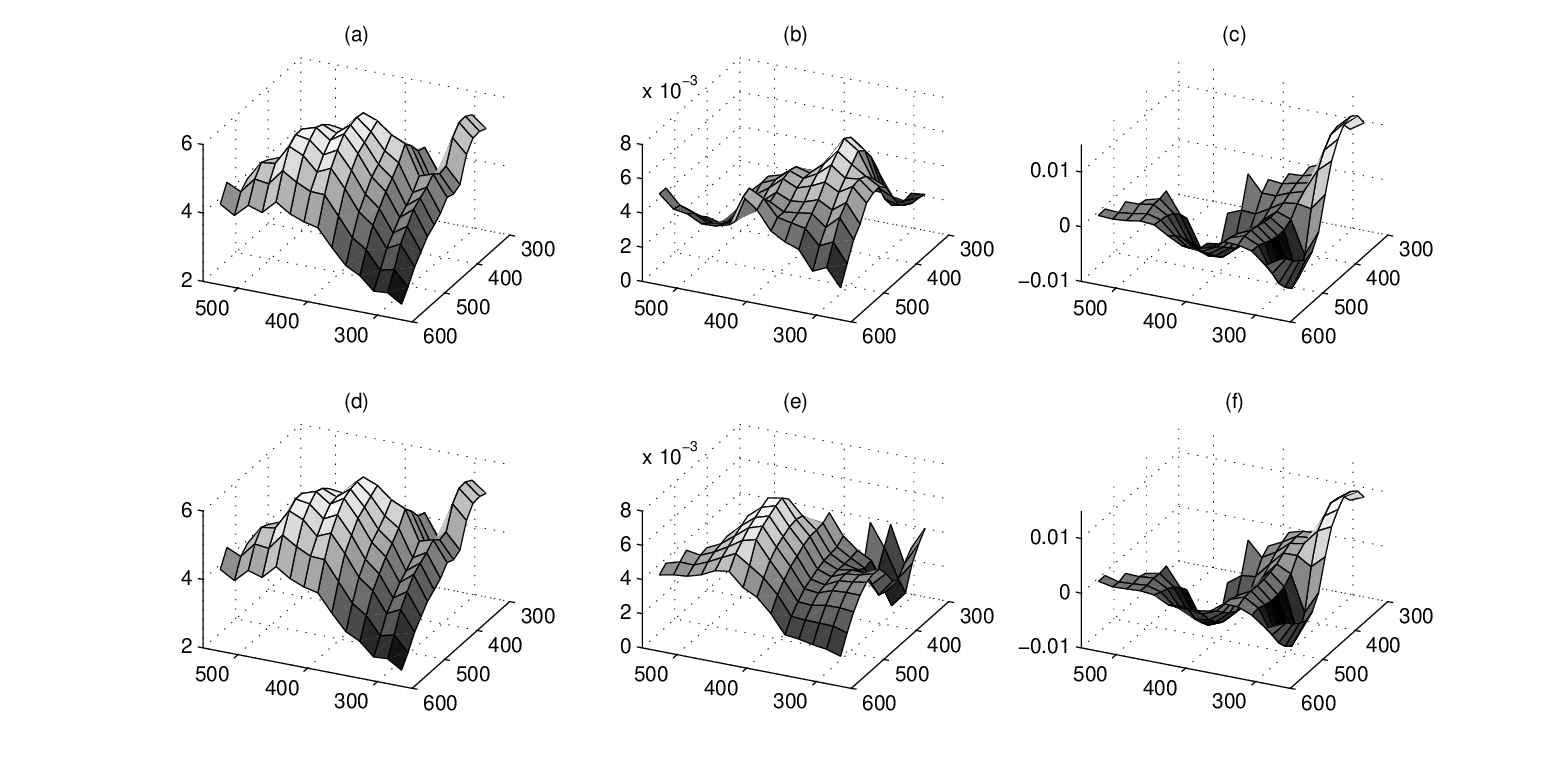';file-properties "XNPEU";}}

While the sample mean (Figure \ref{fig:EEM_mean_pcs}(d)) is not very
different from the trimmed mean (Figure \ref{fig:EEM_mean_pcs}(a)), the
first principal component is seriously affected by the outliers. The first
sample principal component only explains how the outliers vary from the
\textquotedblleft good\textquotedblright\ observations; it may be useful for
outlier detection, but it's not associated with any genuine source of
variability. The first trimmed component, in contrast, is genuinely the main
direction of variability of the \textquotedblleft clean\textquotedblright\
data.

The second trimmed component and the second sample principal component~are
very similar, but the latter underestimates the relative importance of the
component, assigning it only 5\% of the total variability. This is bad
because the second component is the one primarily associated with enzyme
activity. Mortensen and Bro (2006) provide an enzyme activity measure for
calibration, and the correlation coefficient (after eliminating the 15
outliers) between enzyme activity and the second trimmed component is .69.
This association could be overlooked if the user based his analysis on the
non-robust sample principal components and decided that the second component
was negligible.

\subsection{Handwritten Digits}

The planar trajectory of a pen tip is a curve $\mathbf{X}(t)=(x(t),y(t))$ in 
$\mathbb{R}^{2}$, where $t$ is time. Then the analysis of handwritten digits
can be approached as a functional data problem in the Euclidean space $%
\left( \mathcal{L}^{2}(\mathbb{R})\right) ^{2}$ endowed with the inner
product $\langle \mathbf{f},\mathbf{g}\rangle =\int \mathbf{f}(t)^{T}\mathbf{%
g}(t)dt$. The mean trajectory is $\mathbf{\mu }(t)=E\{\mathbf{X}(t)\}$ and
the covariance operator can be represented as 
\begin{equation*}
\mathfrak{C}(\mathbf{f},\mathbf{g})=\diint \mathbf{f}(s)^{T}\mathbf{R}(s,t)%
\mathbf{g}(t)\ ds\ dt
\end{equation*}%
with $\mathbf{R}(s,t)=E[\{\mathbf{X}(s)-\mathbf{\mu }(s)\}\{\mathbf{X}(t)-%
\mathbf{\mu }(t)\}^{T}]$. In this section we analyze a set of 1055
handwritten samples of the digit \textquotedblleft five\textquotedblright ,
available at the Machine Learning Repository of the University of California
at Irvine, \textsf{http://archive.ics.uci.edu/ml/}. The data was rotated and
scaled so that $x$ and $y$ range between 0 and 100, and $t$ between 0 and 1.
Eight sample digits are shown in Figure \ref{fig:digit_samples}.

\FRAME{ftbpFU}{5.9759in}{1.9406in}{0pt}{\Qcb{Handwritten Digits. (a) Sample
mean, (b) 41\% trimmed mean, and (c) mean of the trimmed observations.}}{%
\Qlb{fig:digit_means}}{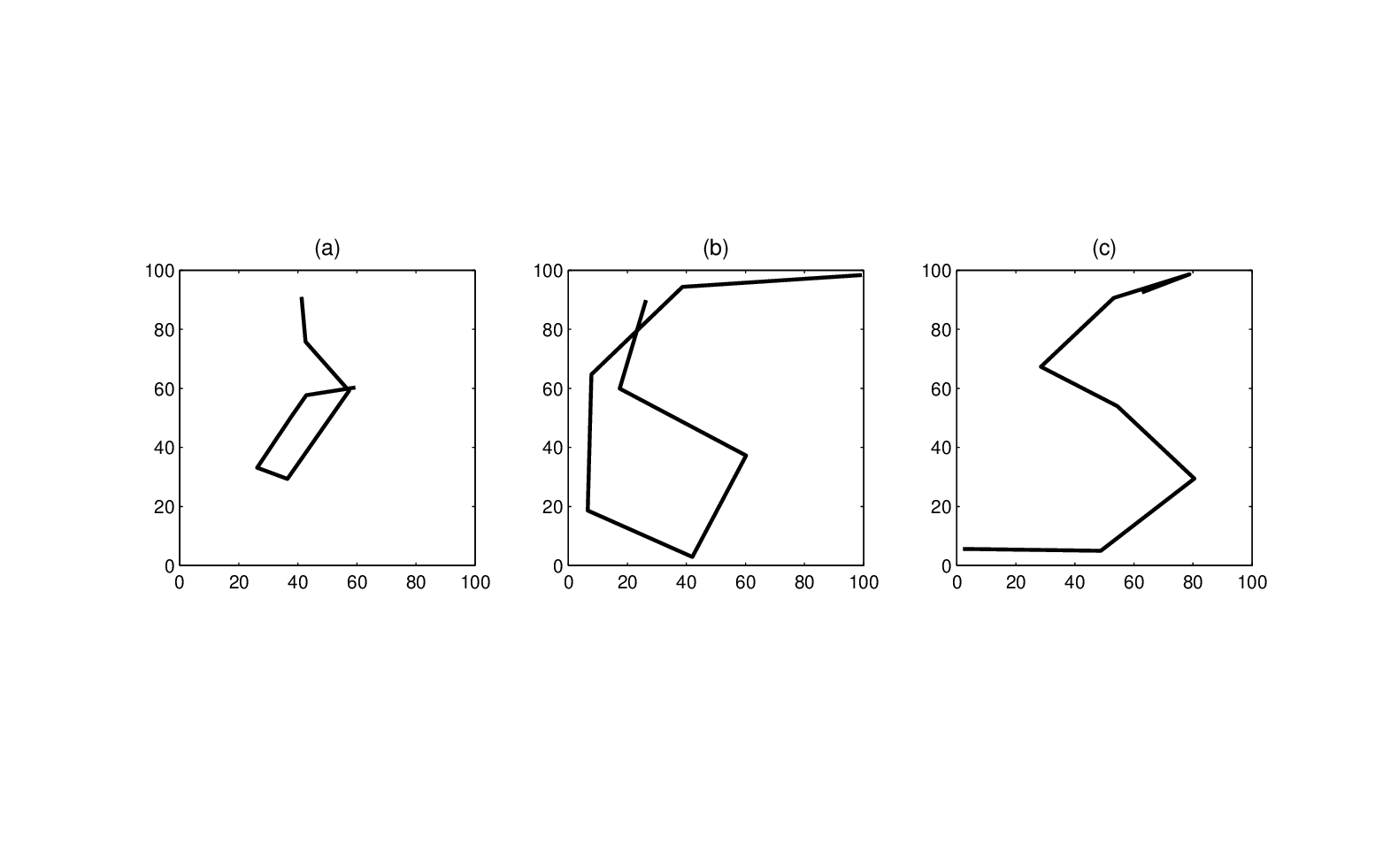}{\special{language "Scientific
Word";type "GRAPHIC";maintain-aspect-ratio TRUE;display "USEDEF";valid_file
"F";width 5.9759in;height 1.9406in;depth 0pt;original-width
10.4184in;original-height 6.5337in;cropleft "0.0480";croptop
"0.7439";cropright "1";cropbottom "0.2556";filename
'digit_means.eps';file-properties "XNPEU";}}

A plot of the sample mean (Figure~\ref{fig:digit_means}(a)) does not
resemble a \textquotedblleft five\textquotedblright\ or any other
recognizable digit. To understand why this happens, we computed the radii
for different values of $\alpha $ and noticed that their distribution
becomes increasingly bimodal as $\alpha $ increases. The histogram for $%
\alpha =.50$ is shown in Figure~\ref{fig:digit_hist_r}. There are two neatly
distinguishable groups: 627 observations with $r_{i}<60$, and 428
observations with $r_{i}>60$. The large number of observations in the second
group (40.5\% of the data) suggests that the sample may be made up of two
systematic clusters, rather than a single homogeneous group and a few
isolated outliers.

This is confirmed by a plot of the 41\% trimmed mean (Figure \ref%
{fig:digit_means}(b)), together with the mean of the observations that were
cut off (Figure \ref{fig:digit_means}(c)). It turns out that there are two
ways to draw the number \textquotedblleft five\textquotedblright . The most
common way is in two strokes, beginning at the upper left corner and moving
downwards, then raising the pen to draw the top dash (but our planar
representation of the trajectory does not capture this vertical movement
explicitly). The other way, less common, is to draw the number
\textquotedblleft five\textquotedblright\ in a single stroke, like the
letter \textquotedblleft S\textquotedblright . Figure \ref{fig:digit_means}%
(b) corresponds to the first class and Figure \ref{fig:digit_means}(c)
corresponds to the second one.

\FRAME{ftbpFU}{3.2301in}{2.4327in}{0pt}{\Qcb{Handwritten Digits. Histogram
of the radii with $\protect\alpha =.50$.}}{\Qlb{fig:digit_hist_r}}{%
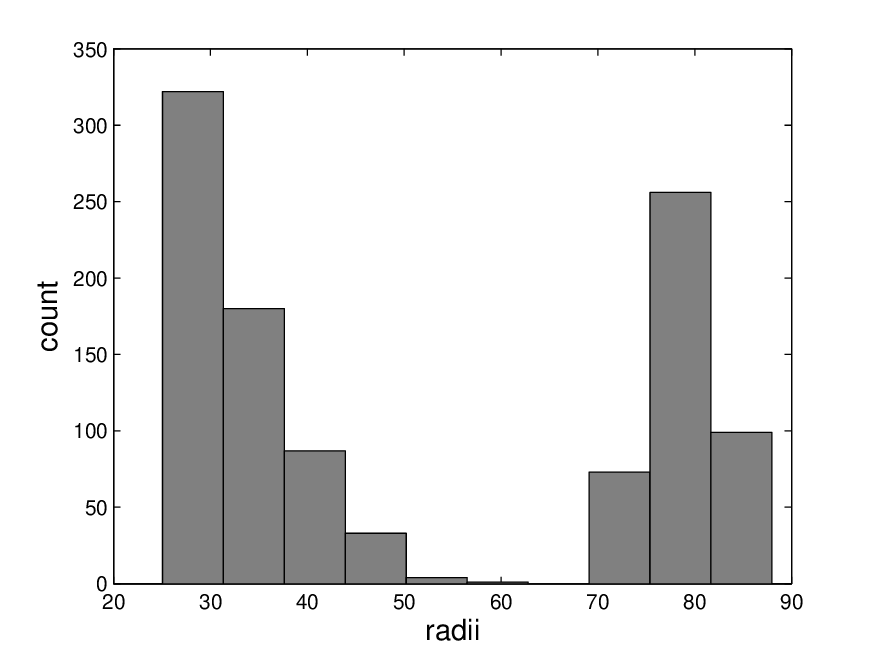}{\special{language "Scientific Word";type
"GRAPHIC";maintain-aspect-ratio TRUE;display "USEDEF";valid_file "F";width
3.2301in;height 2.4327in;depth 0pt;original-width 5.8219in;original-height
4.3708in;cropleft "0";croptop "1";cropright "1";cropbottom "0";filename
'digit_hist_r.eps';file-properties "XNPEU";}}

\FRAME{ftbpFU}{5.2373in}{3.2932in}{0pt}{\Qcb{Handwritten Digits Example.
Effect of the principal components on the mean (----- is the mean; $---$ is
the mean plus 5 times the principal component; $\cdot \cdot \cdot $ is the
mean minus 5 times the principal component). (a) Trimmed mean and first
trimmed component; (b) trimmed mean and second trimmed component; (c) mean
and first component of the trimmed observations; (d) mean and second
component of the trimmed observations.}}{\Qlb{fig:digit_trm_pcs}}{%
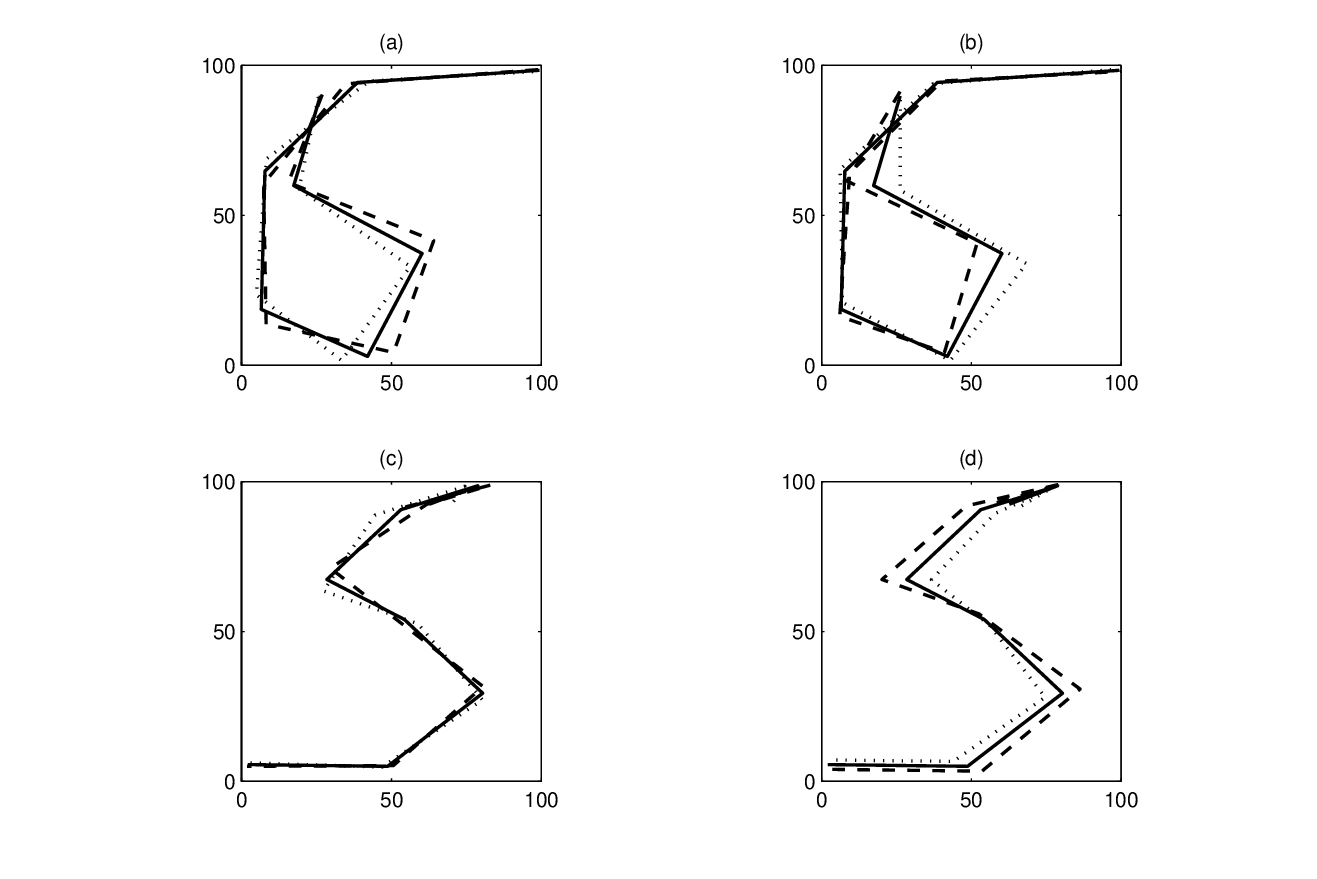}{\special{language "Scientific Word";type
"GRAPHIC";maintain-aspect-ratio TRUE;display "USEDEF";valid_file "F";width
5.2373in;height 3.2932in;depth 0pt;original-width 10.4184in;original-height
6.5337in;cropleft "0";croptop "1";cropright "1";cropbottom "0";filename
'digit_pcs.eps';file-properties "XNPEU";}}

As in the EEMs example, the sample principal components do not provide much
useful information except for discrimination. The trimmed principal
components, on the other hand, do provide useful information about the
directions of variability in the bigger cluster. The easiest way to
interpret the principal components is to plot their effects on the mean
(Figure \ref{fig:digit_trm_pcs}). This figure shows the trimmed mean and the
first two trimmed principal components (Figure \ref{fig:digit_trm_pcs}%
(a,b)), as well as the mean and the first two principal components of the
observations that were cut off (Figure \ref{fig:digit_trm_pcs}(c,d)). The
first trimmed principal component (Figure \ref{fig:digit_trm_pcs}(a))
explains 56\% of the variability and is associated with variation in the
inclination of the \textquotedblleft belly\textquotedblright\ of the digit.
The second trimmed principal component (Figure \ref{fig:digit_trm_pcs}(b))
explains 14\% of the variability and is mostly associated with variation in
the inclination of the vertical dash. Regarding the components of the second
type of \textquotedblleft fives\textquotedblright , the first principal
component (Figure \ref{fig:digit_trm_pcs}(c)) accounts for 44\% of the
variability and is associated with variation in the \textquotedblleft
roundness\textquotedblright\ of the \textquotedblleft five\textquotedblright
: negative scores correspond to rounded \textquotedblleft
S-shaped\textquotedblright\ digits, while positive scores correspond to more
angular \textquotedblleft Z-shaped\textquotedblright\ digits. The second
principal component (Figure \ref{fig:digit_trm_pcs}(d)) accounts for 20\% of
the variability and explains variability in the width of the digit.

\section*{Acknowledgements}

This research was supported by the National Science Foundation, grant DMS
0604396.

\section*{References}

\begin{description}
\item Ash, R.B., and Gardner, M.F. (1975), \emph{Topics in Stochastic
Processes}, Probability and Mathematical Statistics (Vol.~27), New York:
Academic Press.

\item Cuevas, A., Febrero, M., and Fraiman, R. (2007), \textquotedblleft
Robust Estimation and Classification for Functional Data via
Projection-Based Depth Notions,\textquotedblright\ \emph{Computational
Statistics}, 22, 481--496.

\item Donoho, D.L., and Huber, P.J. (1983), \textquotedblleft The Notion of
Breakdown Point,\textquotedblright\ in \emph{A Festschrift for Erich L.
Lehmann}, Belmont, CA: Wadsworth.

\item Fernholz, L. T. (1983), \emph{Von Mises Calculus for Statistical
Functionals}, Lecture Notes in Statistics No.~19, New York: Springer.

\item Filzmoser, P., Serneels, S., Maronna, R., and Van Espen, P.J. (2009),
\textquotedblleft Robust Multivariate Methods in
Chemometrics,\textquotedblright\ in \emph{Comprehensive Chemometrics:
Chemical and Biochemical Data Analysis} (Vol.~III), Amsterdam: Elsevier, pp.
681--722.

\item Fraiman, R., and Meloche, J. (1999), \textquotedblleft Multivariate $L$%
-Estimation,\textquotedblright\ \emph{Test}, 8, 255--317.

\item Fraiman, R., and Muniz, G. (2001), \textquotedblleft Trimmed Means for
Functional Data,\textquotedblright\ \emph{Test}, 10, 419--40.

\item Gervini, D. (2008), \textquotedblleft Robust Functional Estimation
Using the Spatial Median and Spherical Principal
Components,\textquotedblright\ \emph{Biometrika}, 95, 587--600.

\item Gikhman, I.I., and Skorokhod, A.V. (2004), \emph{The Theory of
Stochastic Processes I}, New York: Springer.

\item Gohberg, I., Goldberg, S., and Kaashoek, M.A. (2003), \emph{Basic
Classes of Linear Operators,} Basel: Birkh\"{a}user Verlag.

\item Jiang, J.C., and Mack, Y.P. (2001), \textquotedblleft Robust Local
Polynomial Regression for Dependent Data,\textquotedblright\ \emph{%
Statistica Sinica}, 11, 705--722.

\item Jolliffe, I.T. (2002), \emph{Principal Component Analysis} (2nd~ed.),
Springer Series in Statistics, New York: Springer.

\item Liu, R.Y., Parelius, J.M., and Singh, K. (1999), \textquotedblleft
Multivariate Analysis by Data Depth: Descriptive Statistics, Graphics and
Inference,\textquotedblright\ \emph{The Annals of Statistics}, 27, 783--858.

\item Locantore, N., Marron, J.S., Simpson, D.G., Tripoli, N., Zhang, J.T.,
and Cohen, K.L. (1999), \textquotedblleft Robust Principal Component
Analysis for Functional Data\textquotedblright\ (with discussion), \emph{Test%
}, 8, 1--73.

\item L\'{o}pez-Pintado, S., and Romo, J. (2009), \textquotedblleft On the
Concept of Depth for Functional Data,\textquotedblright\ \emph{Journal of
the American Statistical Association}, 104, 718--734.

\item Lopuha\"{a}, H.P., and Rousseeuw, P.J. (1991), \textquotedblleft
Breakdown Points of Affine Equivariant Estimators of Multivariate Location
and Covariance Matrices,\textquotedblright\ \emph{The Annals of Statistics},
19, 229--248.

\item Maronna, R.A., Martin, R.D., and Yohai, V.J. (2006), \emph{Robust
Statistics. Theory and Methods}, Wiley Series in Probability and Statistics,
New York: Wiley.

\item Mortensen, P.P., and Bro, R. (2006), \textquotedblleft Real-Time
Monitoring and Chemical Profiling of a Cultivation
Process,\textquotedblright\ \emph{Chemometrics and Intelligent Laboratory
Systems}, 84, 106--113.

\item Ramsay, J.O., and Silverman, B.W. (2002), \emph{Applied Functional
Data Analysis. Methods and Case Studies}, Springer Series in Statistics, New
York: Springer.

\item Ramsay, J.O., and Silverman, B.W. (2005), \emph{Functional Data
Analysis }(2nd ed.), Springer Series in Statistics, New York: Springer.

\item Serfling, R. (2006), \textquotedblleft Depth Functions in
Nonparametric Multivariate Inference,\textquotedblright\ in \emph{Data
Depth: Robust Multivariate Analysis, Computational Geometry and Applications
(DIMACS, Vol.~72)}, pp.~1--16.

\item Shi, P.D., and Li, G.Y. (1995), \textquotedblleft Global Convergence
Rates of \emph{B}-spline \emph{M}-estimators in Nonparametric
Regression,\textquotedblright\ \emph{Statistica Sinica}, 5, 303--318.

\item Stigler, S.M. (1973). \textquotedblleft The Asymptotic Distribution of
the Trimmed Mean,\textquotedblright\ \emph{The Annals of Statistics}, 1,
472--477.

\item Van der Vaart, A.W. (1998), \emph{Asymptotic Statistics}, Cambridge
Series in Statistical and Probabilistic Mathematics, Cambridge, UK:
Cambridge University Press.

\item Zuo, Y., and Serfling, R. (2000), \textquotedblleft General Notions of
Statistical Depth Function,\textquotedblright\ \emph{The Annals of Statistics%
}, 28, 461--482.

\item Zuo, Y., Cui, H.,and He, X. (2004), \textquotedblleft On the
Stahel-Donoho Estimator and Depth-Weighted Means of Multivariate
Data,\textquotedblright\ \emph{The Annals of Statistics}, 32, 167--188.
\end{description}

\end{document}